\documentclass[11pt,twoside]{scrartcl}
\KOMAoptions{paper=a4,fontsize=11pt,pagesize=pdftex,headinclude,BCOR=8.5mm,DIV=15}

\usepackage[ngerman]{babel}
\usepackage[utf8]{inputenc}
\usepackage[T1]{fontenc}

\usepackage[slantedGreek]{cmbright}
\usepackage{units}
\usepackage{ellipsis}
\usepackage{amsmath}
\usepackage{endnotes}

\usepackage{tabularx}
\newcolumntype{L}[1]{>{\raggedright\arraybackslash}p{#1}}
\newcolumntype{C}[1]{>{\centering\arraybackslash}p{#1}}
\newcolumntype{R}[1]{>{\raggedleft\arraybackslash}p{#1}}

\usepackage[pdftex]{graphicx}
\usepackage{hyperref}
\hypersetup{
    bookmarks=true,         
    unicode=false,          
    pdftoolbar=true,        
    pdfmenubar=true,        
    pdffitwindow=false,     
    pdftitle={Wo ist Apollo 11?},    
    pdfauthor={Markus Nielbock},     
    pdfsubject={Unterrichtsmaterial},   
    pdfcreator={Markus Nielbock},   
    pdfproducer={},  
    pdfnewwindow=true,      
    colorlinks=true,       
    linkcolor=blue,          
    citecolor=blue,        
    filecolor=blue,      
    urlcolor=blue           
}

\pdftrailerid{} 
\usepackage[european]{circuitikz}
\usepackage[headsepline,footsepline]{scrlayer-scrpage}
\clearscrheadfoot

\usepackage[bf]{caption}
\newcommand\leftmarkline[1]{%
  \parbox[c][\layerheight][b]{\layerwidth}{%
    \hspace*{3.5mm}\rule{#1}{.2mm}%
}}
\DeclareNewLayer[{
  background,
  innermargin,
  oddpage,
  height=.34\paperheight,
  contents={\leftmarkline{2mm}}
}]{oberefaltmarke}
\DeclareNewLayer[{
  clone=oberefaltmarke,
  height=.67\paperheight
}]{unterefaltmarke}
\DeclareNewLayer[{
  clone=oberefaltmarke,
  height=.5\paperheight,
  contents={\leftmarkline{4mm}}
}]{lochermarke}
\AddLayersToPageStyle{@everystyle@}{lochermarke}

\setkomafont{pagefoot}{\normalfont\footnotesize}
\setkomafont{caption}{\normalfont\footnotesize}

\usepackage[babel,german=quotes]{csquotes} 

\usepackage[backend=biber,style=authoryear-icomp]{biblatex}
\newcommand{\myproject}{\Large Unterrichtsmaterialien zur\\ ISS-Horizons-Mission von Alexander Gerst}
\newcommand{\mytitle}{Astronautentraining unter Wasser}

\newcommand{\myclasses}{Klassen 9-11}
\newcommand{\myauthor}{Markus Nielbock}

\rehead*{\small In Kooperation mit\\ \includegraphics[height=11mm]{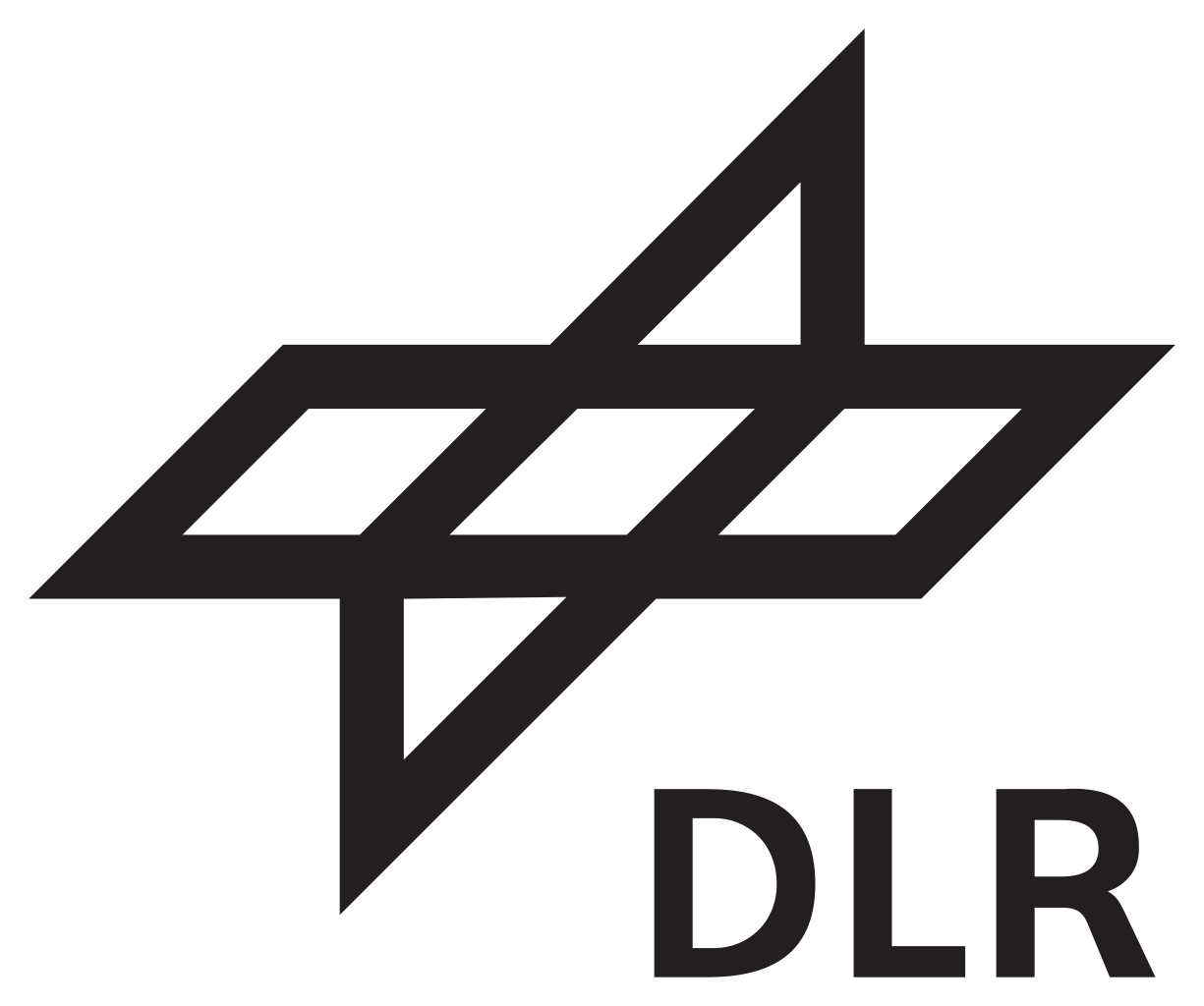}}
\rohead*{\small In Kooperation mit\\ \includegraphics[height=11mm]{DLR_Logo_svg.png}}
\lehead*{\includegraphics[height=15mm]{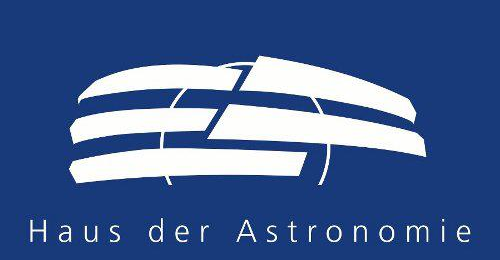}}
\lohead*{\includegraphics[height=15mm]{HdA_500x260.png}}
\ofoot*{Seite \pagemark}
\ifoot*{Handreichung: \mytitle}

\begin{document}

\begin{titlepage}
\thispagestyle{scrheadings}

\begin{center}
{\large\bfseries
Handreichung für Lehrpersonen:

\bigskip
\LARGE
\mytitle
}

\bigskip
{\large\bfseries
\myclasses
}

 \bigskip
{\bfseries
\myauthor
}

\bigskip
2. Mai 2019
\end{center}

\section*{Zusammenfassung}\vspace*{-10pt}
Der Auftrieb sowie das Archimedische Prinzip erscheinen oft als etwas abstrakte Konzepte. Das Unterwasser-Training der ISS-Besatzungen für Außenbordeinsätze bietet einen spannenden Ansatz, um sich mit dem Phänomenen des Auftriebs näher zu beschäftigen. Um einen Schwebezustand im Wasser zu erreichen, müssen die Anzüge mit Gewichten und Auftriebskörpern bestückt werden. Daher ermöglicht diese Aktivität den Schülerinnen und Schülern, durch Aufgaben und Experimente das Wechselspiel zwischen Gewichtskraft und Auftriebskraft zu erforschen.

\section*{Lernziele}\vspace*{-10pt}
Die Schülerinnen und Schüler
\begin{itemize}
\setlength\itemsep{3pt}
\item messen das Gewicht und das Volumen eines Körpers,
\item bringen den Körper im Experiment durch geeignete Maßnahmen unter Wasser zum Schweben,
\item berechnen Gewichts- und Auftriebskraft in verschiedenen Beispielen.
\end{itemize}

\section*{Materialien}\vspace*{-10pt}
\begin{itemize}
\setlength\itemsep{3pt}
\item Arbeitsblätter (erhältlich unter \href{http://www.haus-der-astronomie.de/raum-fuer-bildung}{http://www.haus-der-astronomie.de/raum-fuer-bildung})
\item Stift, Taschenrechner
\item Transparenter Behälter mit Wasser (ca. \unit[15]{$\ell$}, ca. \unit[25]{cm} hoch)
\item Styroporkugel (Durchmesser bis ca. \unit[5]{cm})
\item Haken, möglichst mit Gewinde
\item Gewichte (z. B. Schrauben, Muttern, Büroklammern), Kneifzange oder Seitenschneider (nur unter Aufsicht von Lehrkräften)
\item Digitale Küchenwaage (Genauigkeit ca. \unit[0,1]{g})
\end{itemize}

\section*{Stichworte}\vspace*{-10pt}
Raumstation, ISS, EVA, Raumanzug, Auftrieb, hydrostatischer Druck, Archimedisches Prinzip

\section*{Dauer}\vspace*{-10pt}
120 Minuten

\end{titlepage}

\clearpage



\section*{Hintergrund}
\subsection*{Die Internationale Raumstation}

\begin{figure}[!ht]
\centering
\resizebox{0.6\hsize}{!}{\includegraphics{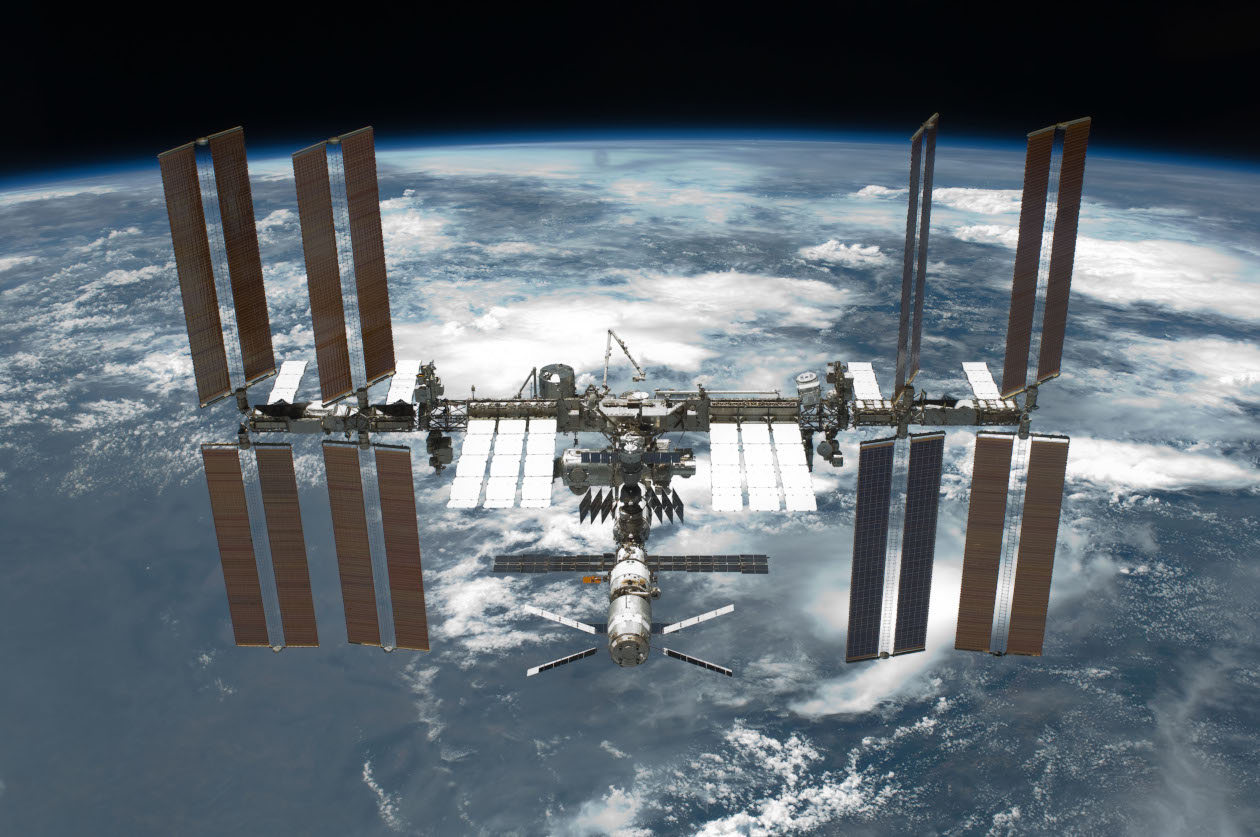}}
\caption{Die ISS im Jahre 2011 (Bild: NASA).}
\label{f:iss}
\end{figure}

\begin{figure}[!ht]
\centering
\resizebox{0.78\hsize}{!}{\includegraphics{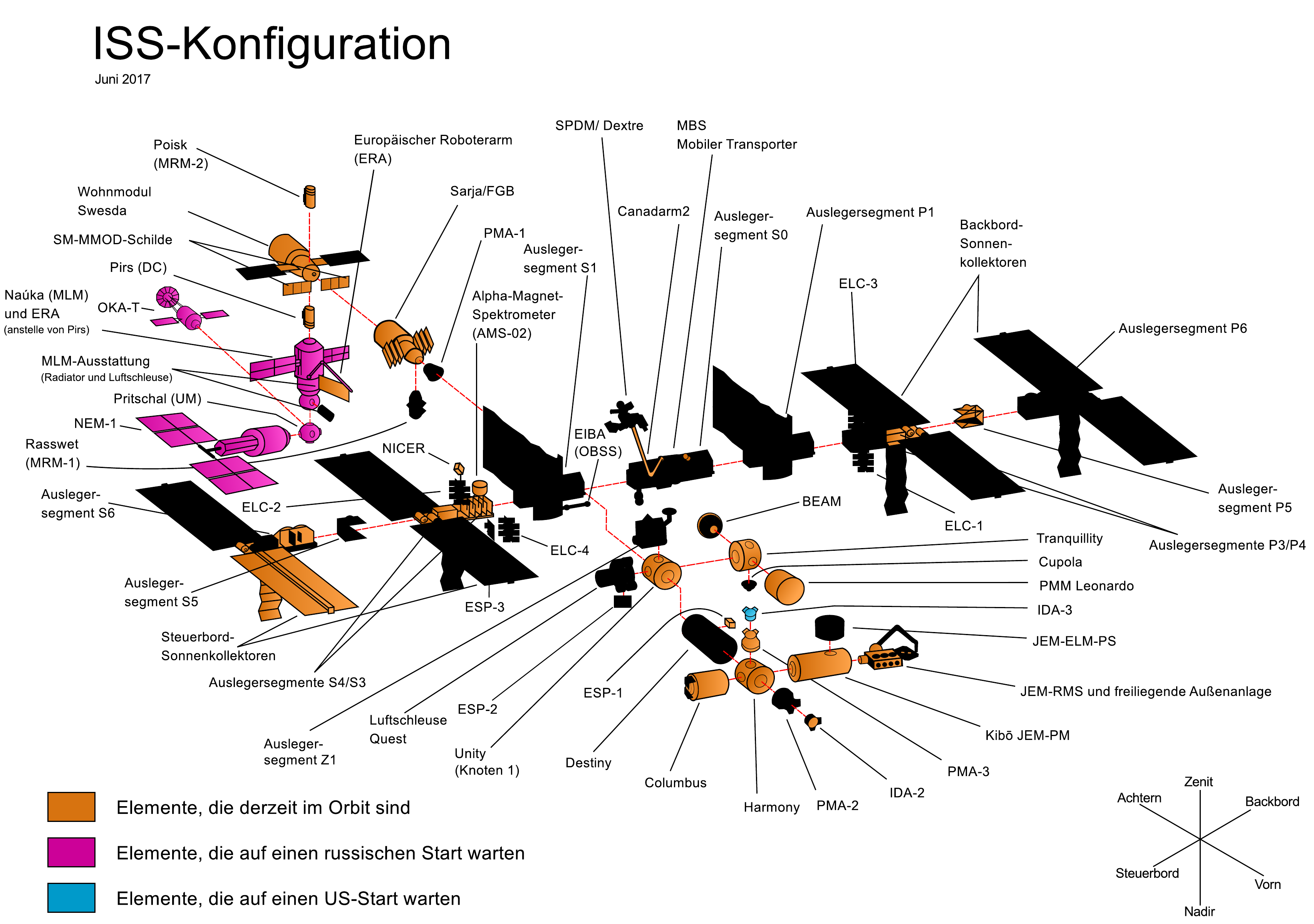}}
\caption{Die Module der ISS im Juni 2017 (Bild: NASA).}
\label{f:modules}
\end{figure}

Seit 1998 wird die Internationale Raumstation (ISS, Abb.~\ref{f:iss}) aufgebaut \autocite{loff_dec._2013} und mittels einzelner Module (Abb.~\ref{f:modules}) ständig erweitert \autocite{zak_after_2017}. Seit 2000 befinden sich in ununterbrochener Reihenfolge Besatzungen an Bord der ISS \autocite{dlr_20_2018}. Ihr Betrieb ist bis mindestens 2024 vorgesehen, wahrscheinlich aber bis 2028 bzw.~2030 möglich \autocite{ulmer_nasa_2015,sputnik_iss_2016,foust_house_2018}. Die gesamte Struktur hat eine Masse von \unit[420]{t}. Sie ist \unit[109]{m} lang, \unit[73]{m} breit \autocite{garcia_international_2016} und \unit[45]{m} hoch \autocite{esa_iss:_2014}. Auf einer Bahnhöhe von etwa \unit[400]{km} benötigt die ISS für eine Erdumrundung ungefähr 92 Minuten \autocite{howell_soyuz_2018}.

Die Internationale Raumstation (Abb.~\ref{f:iss}) ist ein internationales Projekt mit derzeit 15 beteiligten Nationen \autocite{garcia_20_2018,esa_international_2013}. Sie dient als wissenschaftliches Forschungslabor für Fragestellungen, deren experimentelle Untersuchung durch den Einfluss der Gravitation auf der Erde erschwert wird. Man nutzt dafür die an Bord der ISS herrschende Mikrogravitation \autocite{schuttler_mikrogravitationsexperimente_2006}, einen Zustand von annähernder Schwerelosigkeit. Neben der Materialforschung und biologischen Studien spielt auch die Medizin eine wichtige Rolle. Der Einfluss der Mikrogravitation auf die Besatzung führt zu Symptomen, die Krankheitsbildern auf der Erde ähneln. Daher hofft man, Erkenntnisse in der kontrollierten Umgebung der Raumstation zu erlangen, die auch bei der Erforschung der Krankheiten helfen und Therapien den Weg ebnen \autocite{buhrke_was_2018}. Ein weiterer Grund besteht darin, langfristige Missionen innerhalb des Sonnensystems vorzubereiten \autocite{ganse_weltraummedizin:_2018}.

\medskip
Manchmal sind Außenbordeinsätze (EVA, Extra-vehicular Activity) erforderlich, um Reparaturen vorzunehmen oder neue Ausrüstung zu montieren. Wie sich die Astronauten darauf vorbereiten, wird im folgenden Kapitel beschrieben.

\subsection*{Astronautentraining}
Arbeiten außerhalb der schützenden Hülle der ISS sind immer wieder notwendig. Insbesondere der Aufbau der Raumstation war durch zahlreiche EVA gekennzeichnet. Aber auch heute müssen immer wieder Reparaturen durchgeführt oder neue Ausrüstung montiert werden\endnote{Eine Liste aller EVA findet man unter:\\ \url{https://de.wikipedia.org/wiki/Liste\_der\_Weltraumausstiege}}. Für die Ausstiege werden Raumanzüge benötigt, die der Besatzung einen ausreichenden Schutz gegen das Vakuum und die Strahlung im Weltall bieten. Handgriffe, Bewegungen und das zielgerichtete Ansteuern sind mit dieser Hülle schwierig und müssen erlernt werden.

\begin{figure}[!ht]
\centering
\resizebox{0.75\hsize}{!}{\includegraphics{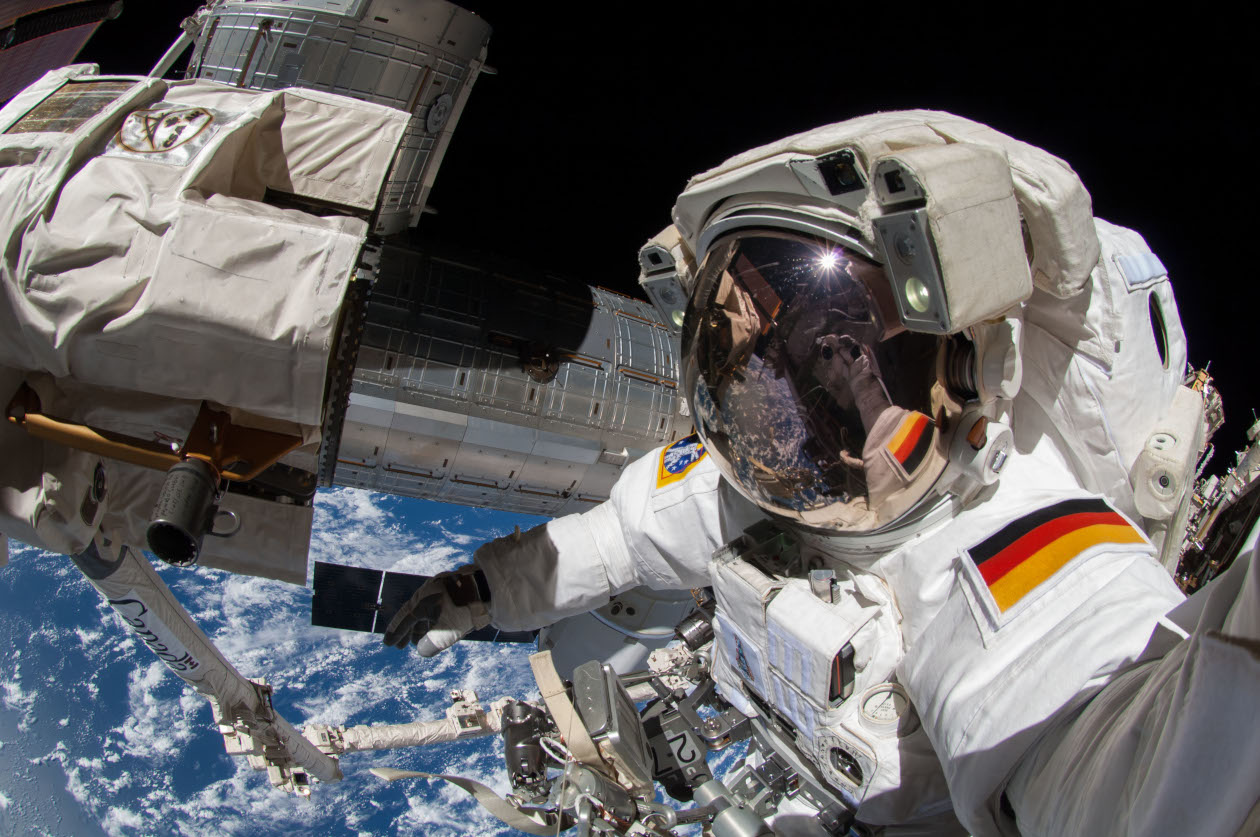}}
\caption{Außenbordeinsatz auf der ISS von Alexander Gerst im Jahr 2014 (Bild: NASA).}
\label{f:eva_gerst}
\end{figure}

Um die Arbeiten in den Raumanzügen routiniert und sicher ausführen zu können, werden die Aktivitäten auf der Erde in Wasserbecken trainiert \autocite{esa_spacewalk_2013,gast_glimpse_2009}. Indem man die Astronauten mit ihren Raumanzügen im Wasser zum Schweben bringt, erzeugt man einen Zustand der dem der Schwerelosigkeit auf der ISS recht nahe kommt. Allerdings erzeugt das Wasser einen Widerstand, den man im All nicht spürt. Zudem sind die Astronauten in den Anzügen während des Trainings unter Wasser nicht schwerelos.

\medskip
Der Raumanzug (EMU, Extra-vehicular Moblity Unit) wiegt inklusive Astronaut ca.~\unit[200]{kg} \autocite{tate_how_2013,thomas_u._2011}. Um dem Anzug die nötige Balance aus Auftrieb und Gewicht zu verleihen, wird er mit Schwimmkörpern und Gewichten versehen. Zudem wird der Luftdruck innerhalb der EMU während des Trainings stets auf \unit[0,3]{bar} (\unit[4,3]{PSI}) über dem Umgebungsdruck gehalten \autocite{hutchinson_swimming_2013}. Dadurch bläht sich der Anzug leicht auf, was ebenfalls zu einem Auftrieb führt. Gleichzeitig wird darauf geachtet, dass während des Trainings Auftrieb und Gewicht so ausbalanciert werden, dass der gesamte Körper in einem neutralen Gleichgewicht ist. Ansonsten könnte es sein, dass sich der Astronaut entweder ständig aufrichtet oder kopfüber im Wasser schwimmt.

\begin{figure}[!ht]
\centering
\resizebox{0.88\hsize}{!}{\includegraphics{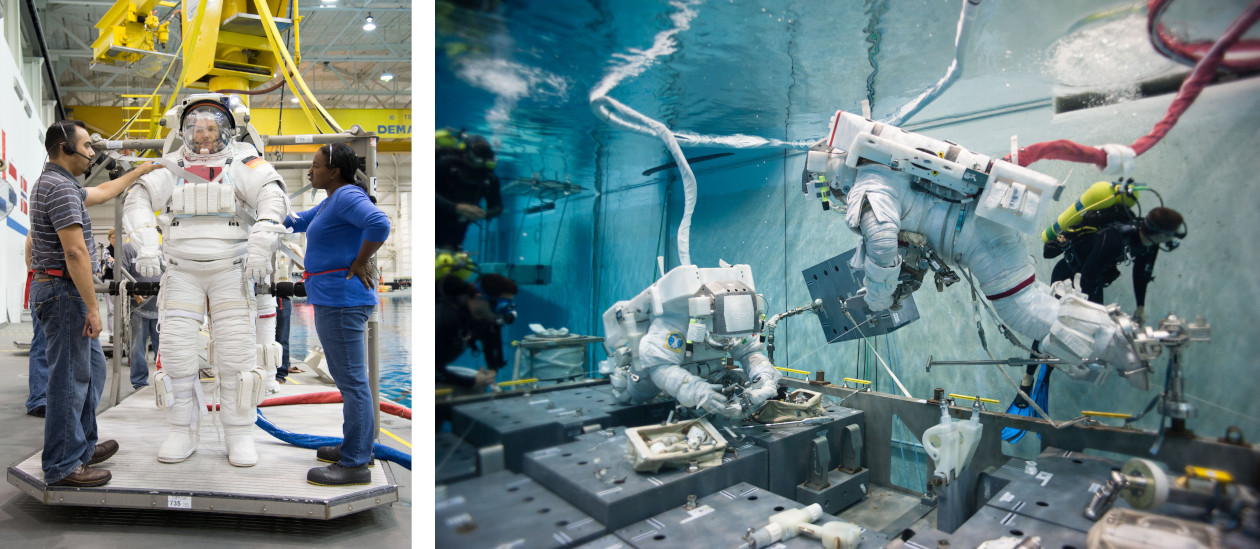}}
\caption{Links: Alexander Gerst wird für das Training im Wasserbecken vorbereitet. Rechts: Training von Arbeitsschritten unter Wasser. (Bilder: NASA).}
\label{f:nbl1}
\end{figure}

Für die europäischen ESA-Astronauten stehen dafür die Neutral Buoyancy Facility am Europäischen Astronautenzentrum (EAC) in Köln sowie das Neutral Buoyancy Lab der NASA an der Sonny Carter Training Facility nahe des Johnson Space Center in Houston, Texas zur Verfügung.

\begin{figure}[!ht]
\centering
\resizebox{0.88\hsize}{!}{\includegraphics{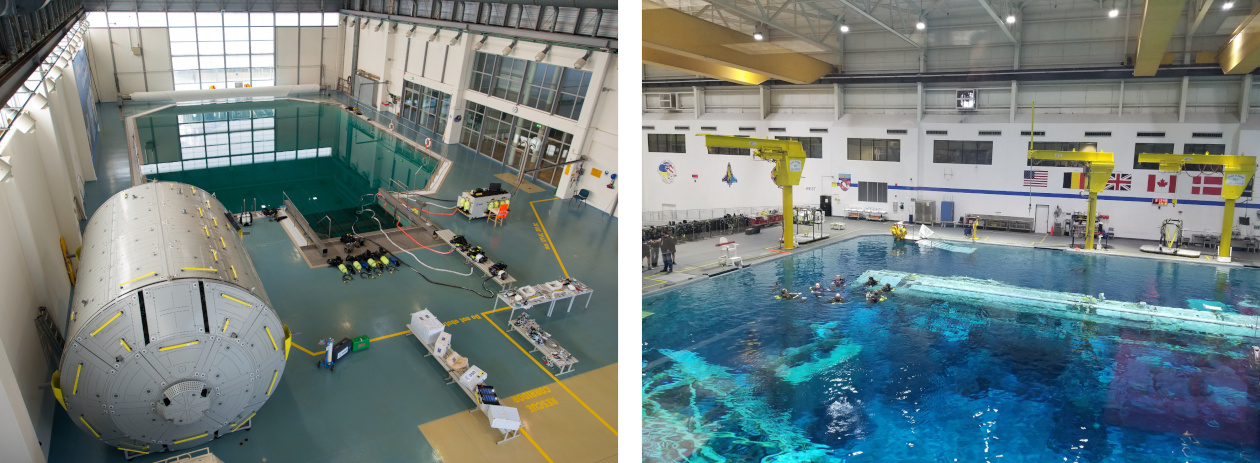}}
\caption{Links: das Becken der Neutral Buoyancy Facility der ESA am EAC in Köln (Bild: ESA–S. Corvaja, 2015). Rechts: das Neutral Buoyancy Lab der NASA mit Aufbauten von ISS-Strukturen unter Wasser (Bild: CCicalese (WMF), CC BY 4.0).}
\label{f:nbl2}
\end{figure}

Mit den Planungen für eine Rückkehr zum Mond wird man diese Trainingsmethode auch zur Simulation der verminderten Schwerkraft auf dem Mond nutzen. Studien wie Moondive \autocite{esa_mondspaziergange_2018} ermitteln bereits jetzt, welche Anforderungen für die spezielle Umgebung des Monds erfüllt sein müssen und ob die bereits vorhandenen Trainingseinrichtungen entsprechend angepasst werden können. Nachfolgend wird erläutert, welche physikalischen Kräfte für den Auftrieb und das Schweben unter Wasser verantwortlich sind.

\subsection*{Der hydrostatische Druck}
\begin{figure}[!ht]
\centering
\resizebox{0.4\hsize}{!}{\includegraphics{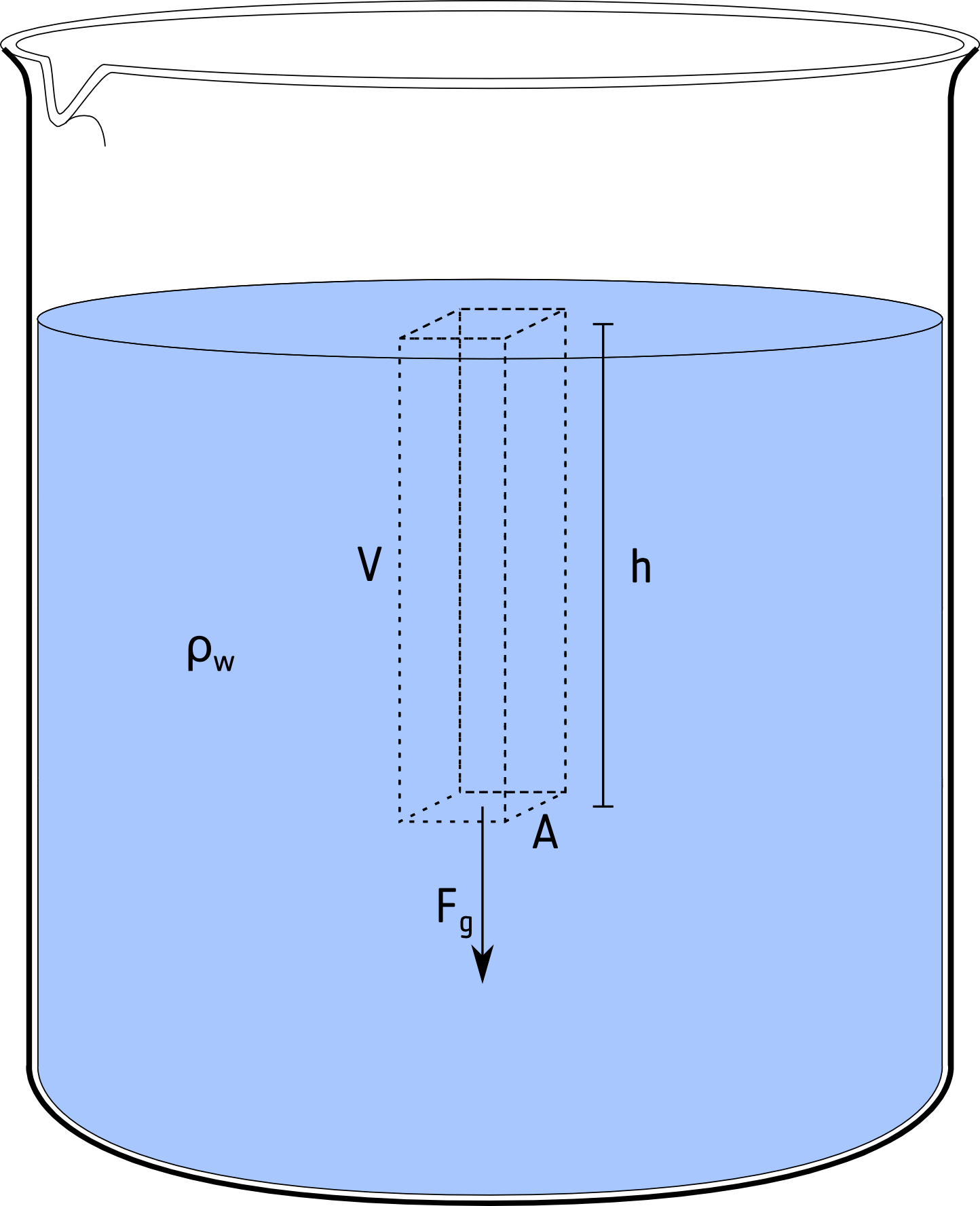}}
\caption{Die Grafik verdeutlicht die Verhältnisse, die zum hydrostatischen Druck bzw. Schweredruck führen (Grafik: eigenes Werk).}
\label{f:druck}
\end{figure}

Um zu verstehen, wodurch der Auftrieb in einem Medium wie dem Wasser erzeugt wird, benötigen wir den Begriff des hydrostatischen Drucks. Dabei handelt es um den Druck, der in einer bestimmten Tiefe des entsprechenden Mediums durch die Gewichtskraft der darüber liegenden Säule erzeugt wird. Zur Verdeutlichung soll Abb.~\ref{f:druck} betrachtet werden.

Dargestellt ist ein Becherglas mit Wasser. In einer Tiefe $h$ befinde sich eine gedachte Fläche $A$, die zusammen mit $h$ das Volumen $V$ einer Wassersäule (gestrichelt) bildet. Dieses Wasser mit der Dichte $\varrho_w$ übt auf $A$ die Gewichtskraft $F_g$ aus. Die Masse des Wassersäule beträgt $m_w = \varrho_w \cdot V$. Daraus folgt:

\begin{align}
F_g &= m_w\cdot g \notag \\[5pt]
    &= \varrho_w \cdot V \cdot g \\[5pt]
    &= \varrho_w \cdot h \cdot A \cdot g \notag \\[5pt]
\Leftrightarrow p &= \frac{F_g}{A} = \varrho_w \cdot h \cdot g
\end{align}

\medskip
Somit steigt der Druck $p$ proportional mit der Wassertiefe $h$. Tatsächlich ergibt sich der Gesamtdruck aus dem Druck der Wassersäule und dem Luftdruck der darüber liegenden Atmosphäre, bezeichnet mit $p_0$. Somit kann der hydrostatische Druck insgesamt formuliert werden als:

\begin{equation}
p = \varrho \cdot g \cdot h + p_0
\label{e:druck}
\end{equation}

\subsection*{Der Auftrieb}
\begin{figure}[!ht]
\centering
\resizebox{0.4\hsize}{!}{\includegraphics{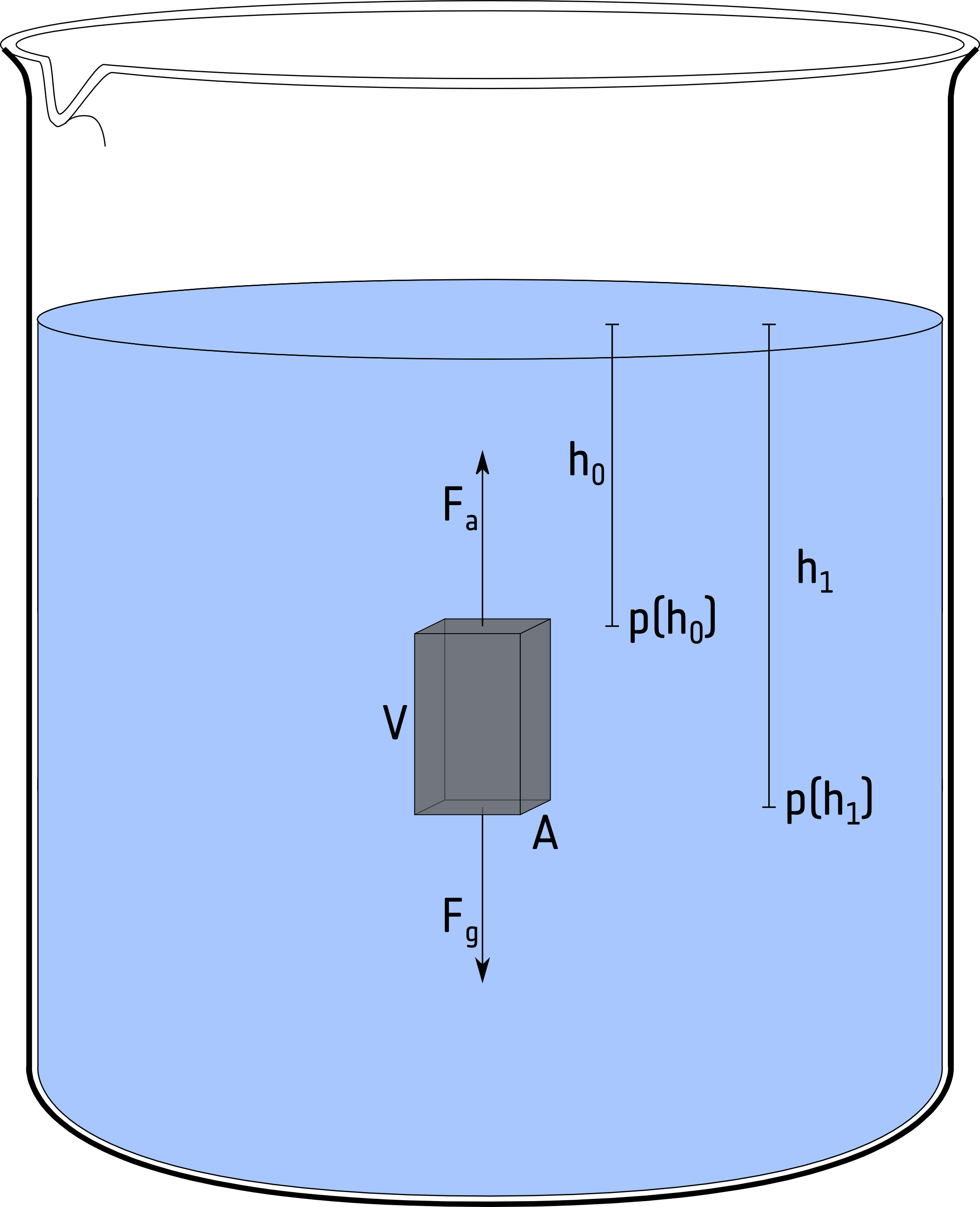}}
\caption{Die Grafik verdeutlicht das Phänomen des Auftriebs, den ein Körper mit dem Volumen $V$ unter Wasser erfährt (Grafik: eigenes Werk).}
\label{f:auftrieb}
\end{figure}

Ein Objekt mit der Masse $m$ wird im Schwerefeld der Erde mit dem Ortsfaktor $g$ zum Erdmittelpunkt beschleunigt und erfährt eine Gewichtskraft $F_g = m\cdot g$. Taucht man jedoch dieses Objekt in ein Gefäß mit Wasser, wird die Wirkung von $F_g$ um die Auftriebskraft $F_a$ reduziert. Abbildung~\ref{f:auftrieb} verdeutlicht die Situation. Sie zeigt einen Körper, der ein Volumen $V$ und eine Grundfläche $A$ aufweist. Er habe die Masse $m$. Die Oberkante der Körpers hat die Tiefe $h_0$, die Unterkante die Tiefe $h_1$. Somit besitzt das Objekt selbst die Höhe $\Delta h = h_1 - h_0$ mit $V = A\cdot \Delta h$. Nun lässt sich der hydrostatische Druck $p$ bei $h_0$ und $h_1$ ermitteln.
{
\setlength{\abovedisplayskip}{0pt}
\setlength{\abovedisplayshortskip}{0pt}
\setlength{\belowdisplayskip}{0pt}
\setlength{\belowdisplayshortskip}{0pt}
\begin{align}
    p(h_0) &= \varrho_w \cdot g \cdot h_0 + p_0\notag \\[5pt]
    p(h_1) &= \varrho_w \cdot g \cdot h_1 + p_0\notag
\end{align}
}

Da $h_1 > h_0$, folgt $p(h_1) > p(h_0)$. Damit ist die auf $A$ wirkende Kraft $F = p\cdot A$ bei $h_1$ ebenfalls größer als bei $h_0$. Somit überwiegt die auf die Unterseite des Körpers wirkende Kraft derjenigen auf seiner Oberseite. Die daraus resultierende Kraftdifferenz wirkt daher nach oben, also entgegen der Richtung der Gravitation. Dies ist die Auftriebskraft $F_a$.

\begin{align}
    F_a &= \Delta p \cdot A = \left(p(h_1)-p(h_0)\right) \cdot A \notag \\[5pt]
        &= \left(\varrho_w \cdot g \cdot h_1 + p_0 - \left(\varrho_w \cdot g \cdot h_0 + p_0\right)\right) \cdot A \notag \\[5pt]
        &= \varrho_w \cdot g \cdot \left(h_1 - h_0\right) \cdot A \notag \\[5pt]
        &= \varrho_w \cdot g \cdot \Delta h \cdot A \notag \\[5pt]
        &= \varrho_w \cdot g \cdot V \\[5pt]
        &= m_w \cdot g
\end{align}

\medskip
Daraus folgt das \textbf{Archimedische Prinzip}\endnote{Archimedes von Syrakus (ca. 287 bis 212 v.~d.~Z.) war ein griechischer Mathematiker, Physiker, Ingenieur, Erfinder und Astronom.}, das besagt:

\begin{quote}
"`Die Auftriebskraft entspricht der Gewichtskraft des Mediums, das von dem eingetauchten Volumen verdrängt wird."'
\end{quote}

Daraus lassen sich drei Fälle ableiten.

\begin{itemize}
    \item $F_g > F_a$: Der Körper sinkt.
    \item $F_g < F_a$: Der Körper steigt.
    \item $F_g = F_a$: Der Körper schwebt im Wasser bzw. schwimmt an der Oberfläche.
\end{itemize}

Die letztere Situation des Kräftegleichgewichts macht man sich im Astronautentraining zunutze. Bei diesem Gleichgewicht schwebt der Astronaut im Wasserbecken. Doch wie führt man dieses Kräftegleichgewicht herbei? Welche Größe ist hierbei maßgeblich? Eine einfache Rechnung ergibt die Antwort.

\begin{align}
    F_g &= F_a \label{e:Fgg} \\[5pt]
    \Leftrightarrow m\cdot g &= m_w \cdot g \notag \\[5pt]
    \Leftrightarrow \varrho_k \cdot V \cdot g &= \varrho_w \cdot V \cdot g \notag \\[5pt]
    \Leftrightarrow \varrho_k &= \varrho_w
\end{align}

\medskip
Der Körper schwebt im Wasser, wenn seine Dichte $\varrho_k$ der des Wassers entspricht. Sinkt er, ist seine Dichte zu hoch. Um die Dichte zu verringern, muss man bei gleichbleibender Masse das Volumen erhöhen. Das geschieht im Fall der Astronauten im Trainingsbecken durch Luftpolster. Sie erhöhen das Volumen stärker als im Vergleich die Masse, die sie mit sich tragen. In der Bilanz wird also die mittlere Dichte des Astronauten verringert. Das führt zu einem verstärkten Auftrieb. Nun kann man so lange Luftpolster anbringen, bis im Mittel die Dichte von Wasser erreicht ist. Der Astronaut schwebt\endnote{Stattdessen könnte man theoretisch auch die Dichte des Mediums verändern. Salzwasser hat eine höhere Dichte als Süßwasser.}.

\clearpage
\section*{Aktivität: }

\subsection*{Vorbereitung für Lehrpersonen}
Beschaffen Sie die auf den Deckblatt angegebenen Materialien in einer Stückzahl, die für Ihre Lerngruppe angemessen ist. Beachten Sie, dass Sie für die Durchführung des Experiments Wasser benötigen.

\medskip
Bedenken Sie, dass die Experimentierzutaten lediglich als Beispiel anzusehen sind, die der Autor getestet hat. Allerdings können beispielsweise auch kleinere Styroporkugeln verwendet werden, die einen entsprechend geringeren Auftrieb erzeugen. Größere Kugeln sind nicht zu empfehlen, da hierfür deutlich schwerere Gewichte und somit auch ggf.~größere Wasserbehälter benötigt werden. Es bietet sich an, vorab das Experiment selbst zu testen, um den Umfang der erforderlichen Gewichte zu ermitteln.

\medskip
Das Experiment darf nur unter Aufsicht einer Lehrkraft durchgeführt werden. Der Autor und die Projektpartner dieses Materials übernehmen keine Haftung für etwaig auftretende Verletzungen.

\medskip
Die in der Einführung angegebenen Videos (s.~u.) benötigen einen Internetzugang. Falls Sie in der Lehreinrichtung nicht über die notwendige Ausstattung verfügen, weisen Sie die Schülerinnen und Schüler vorab an, sich diese Videos als Vorbereitung zuhause anzusehen. Ermutigen Sie sie, sich davon ausgehend weiter über das Thema zu informieren. Eine gute Seite zur Eigenrecherche ist:

\medskip
DLR\_next -- Raumfahrt\\
\url{https://www.dlr.de/next/desktopdefault.aspx/tabid-6100/}

\subsection*{Thematische Einführung (Vorschlag)}

Fragen Sie die Schülerinnen und Schüler, ob sie wissen, was ein Weltraumspaziergang ist. Zeigen Sie zur Erläuterung das folgende Video.

\bigskip
DLR\_next -- Mit Alex ins All: Der Spacewalk (Dauer: 5:04 min)\\
\url{https://youtu.be/vZUAFdJ3C-4}

\bigskip
Die im Video erwähnten Liveübertragungen der NASA kann man über die folgende Webadresse sehen:\\
\url{https://www.nasa.gov/nasalive}

\bigskip
Bei weiterem Interesse können sich die Schülerinnen und Schüler die folgenden beiden Videos ansehen. Da sie in englischer Sprache produziert sind, ermöglichen Sie zudem eine Verknüpfung mit dem Fach Englisch und schulen das Verständnis gesprochener Sprache.

\bigskip
Alexander Gerst training in Houston (Dauer: 3:26 min, englisch)\\
\url{https://youtu.be/f4cHgAIK79c}

\bigskip
How Astronauts Train Underwater at NASA's Neutral Buoyancy Lab (Dauer: 7:06 min, englisch)\\
\url{https://youtu.be/BRPb0J8lZcY}

\bigskip
Fragen Sie, wovon es abhängt ob ein Objekt im Wasser schwimmt oder untergeht. Wahrscheinlich wird hier der Begriff Gewicht genannt. Als Gegenbeispiel können Sie den Vergleich eines Steins mit einem riesigen Containerschiff aufstellen. Damit ist das Gewicht bzw.~die Masse offenbar nicht die bestimmende Größe. Offenbar ist das Verhältnis von Masse und Volumen maßgeblich, oder in anderen Worten: die Dichte.

\medskip
Schließlich schauen sich die Schülerinnen und Schüler zur Einführung oder zur Auffrischung des Prinzips des Auftriebs die folgenden Videos an.

\bigskip
Archimedes von Syrakus - Gold oder nicht Gold? (Dauer: 3:24 min)\\
\url{https://youtu.be/e2hcOCSd-xw}

\bigskip
Archimedisches Prinzip – Der Auftrieb (Dauer: 4:45 min)\\
\url{https://youtu.be/ZUWrj-nqkE4}

\bigskip
Auftrieb (Dauer: 1:03 min)\\
\url{https://youtu.be/FPTGbb9p0po}

\bigskip
Fragen Sie zum Schluss, worauf man achten muss, damit die Astronauten während des Trainings im Wasserbecken schweben. Eine korrekte Antwort sollte beinhalten, dass das sich Gewicht und Auftrieb möglichst genau ausgleichen müssen. Ein Objekt schwebt dann im Wasser, wenn es im Mittel dieselbe Dichte wie Wasser hat.

\clearpage
\subsection*{Experiment}
{\em Hinweis! Der Inhalt des entsprechenden Kapitels im Arbeitsmaterial für die Schülerinnen und Schüler ist gegenüber dieser Version reduziert. Einzelne Zwischenschritte und Lösungen sind aus didaktischen Gründen dort nicht enthalten.}

\medskip
Das Experiment stellt im Modell nach, was es bedeutet, ein Objekt (z.~B.~eine Astronautin oder einen Astronauten) im Wasser zum Schweben zu bringen. Eine Styroporkugel dient hierbei als Objekt mit geringer Dichte, das auf der Wasseroberfläche schwimmt. Geeignete Gewichte müssen so lange an einen daran befestigten Haken angehängt werden, bis die Kugel im Wasser schwebt. Dabei wird man feststellen, dass es sehr schwierig ist, Auftrieb und Gewicht exakt zum Ausgleich zu bringen. In dem dargestellten Beispiel werden Schrauben, Muttern und Büroklammern als Gewichte benutzt.

\begin{figure}[!ht]
\centering
\resizebox{0.5\hsize}{!}{\includegraphics{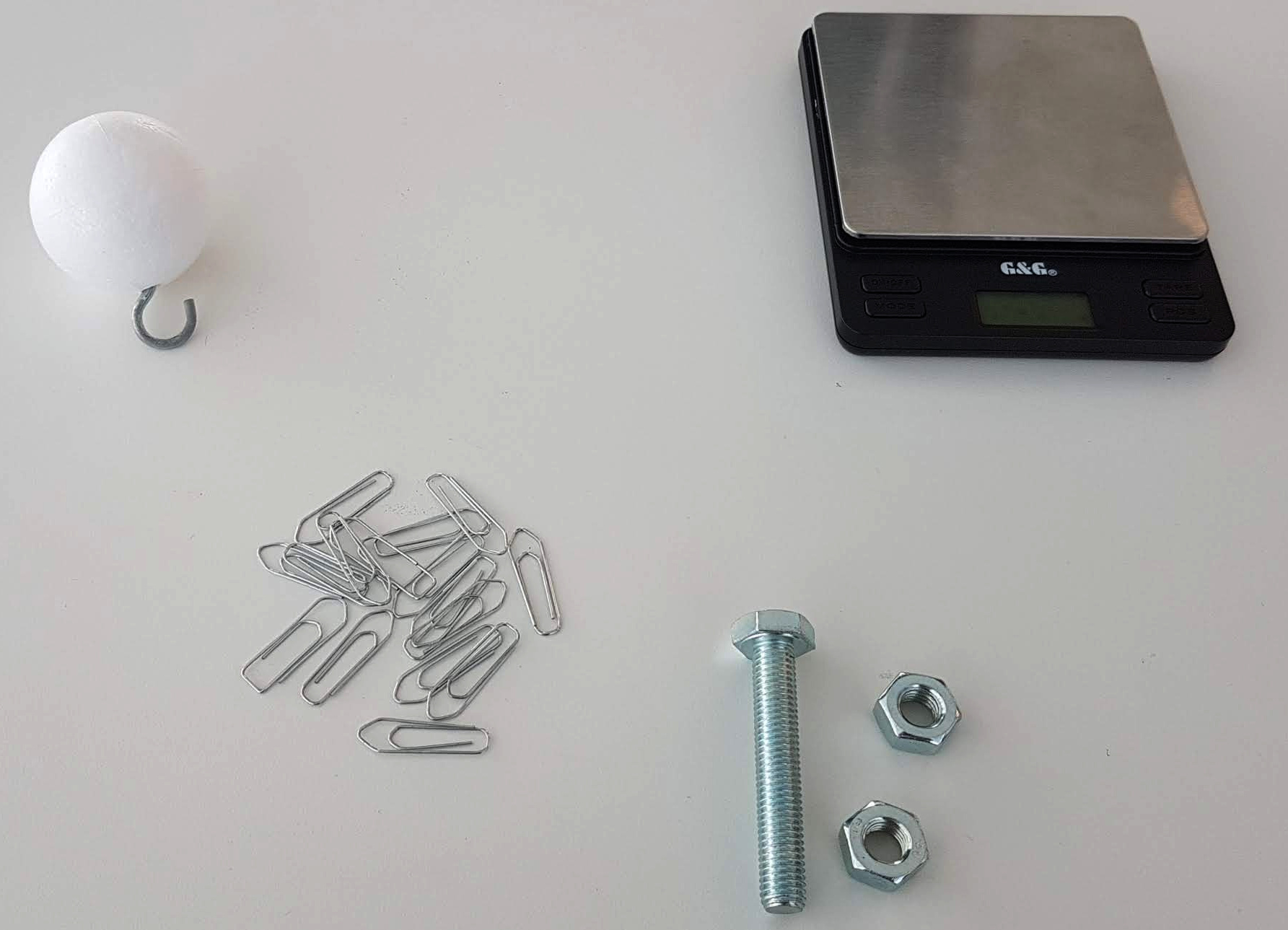}}
\caption{Eine Styroporkugel und Gewichte. Ihr Gesamtgewicht -- gewogen durch eine Digitalwaage -- soll im Experiment den Auftrieb im Wasser ausgleichen (Bild: M. Nielbock).}
\label{f:exp_aufbau}
\end{figure}

Zu Beginn schwimmt die Kugel auf dem Wasser.

\begin{figure}[!ht]
\centering
\resizebox{0.5\hsize}{!}{\includegraphics{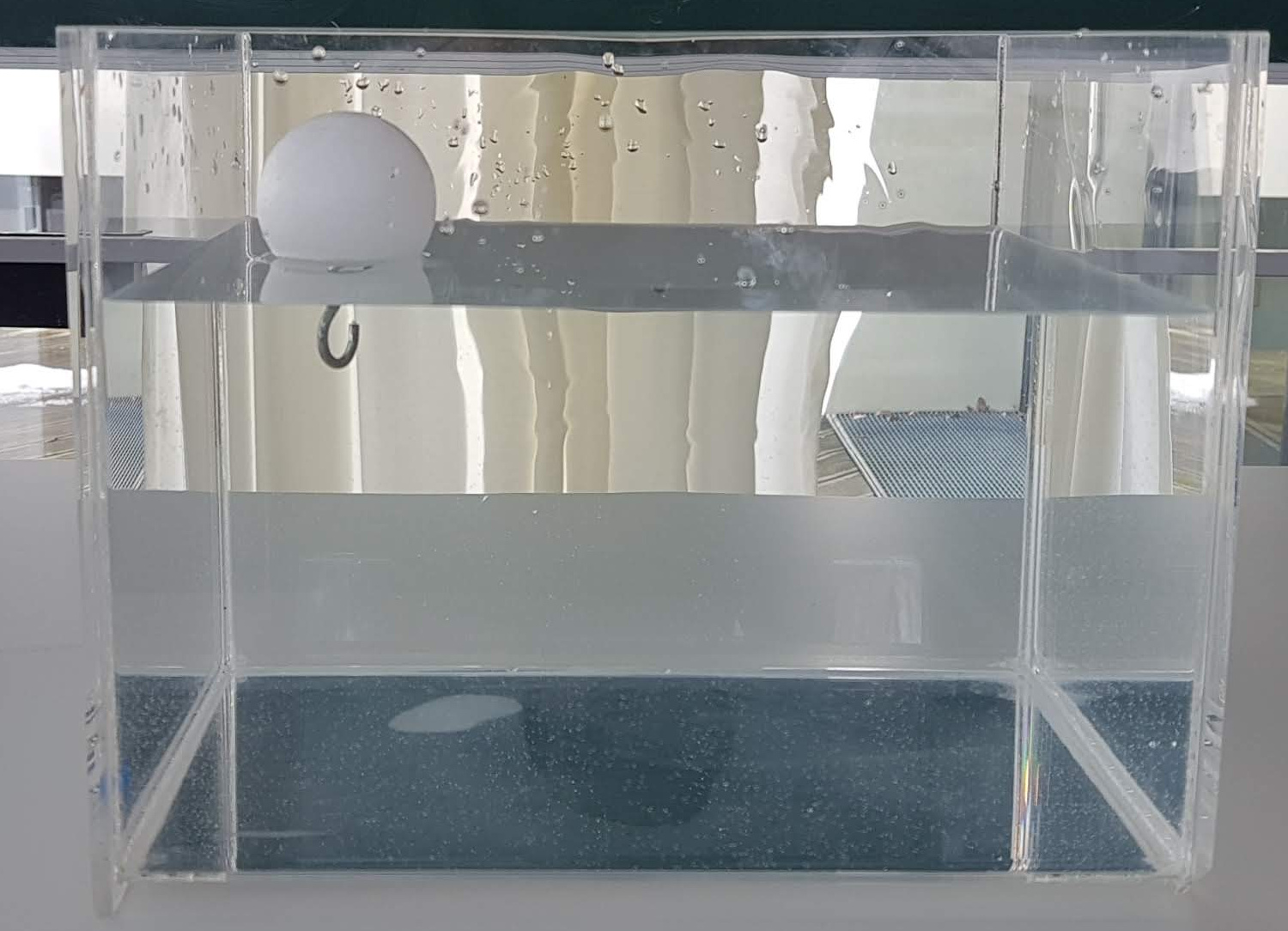}}
\caption{Die Kugel schwimmt auf dem Wasser (Bild: M. Nielbock).}
\label{f:exp_step1}
\end{figure}

\subsubsection*{Erläutere, wie man die Masse abschätzen kann, die benötigt wird, damit die Styroporkugel im Wasser schwebt.}

Das Archimedische Prinzip besagt, dass die Kraft des Auftriebs der Gewichtskraft des Wassers entspricht, das durch das Volumen des eingetauchten Körpers verdrängt wird. Daher kann man als erste Abschätzung das Gewicht bzw.~die Masse des Wassers innerhalb eines Volumens berechnen, das dem Volumen der Styroporkugel entspricht.

\begin{table}[!ht]
    \centering
    \renewcommand{\arraystretch}{1.3}
    \caption{Wichtige physikalische Größen und ihre Werte.}
    \label{t:groessen}
    \begin{tabular}{lcc}
    \hline
    Größe & Formelzeichen & Zahlenwert und Einheit \\
    \hline
    Erdbeschleunigung & $g$ & \unit[9,81]{m/s$^2$} \\
    Atmosphärischer Normaldruck & $p_0$ & \unit[1013,25]{hPa} $=$ \unit[101325]{Pa} \\
    Dichte von Wasser & $\varrho_w$ & \unit[997]{kg/m$^3$} \\
    Dichte von Gold & $\varrho_\mathsf{Au}$ & \unit[1939]{kg/m$^3$} \\
    Dichte von Silber & $\varrho_\mathsf{Ag}$ & \unit[1049]{kg/m$^3$} \\
    \hline
    \end{tabular}
\end{table}

\begin{figure}[!ht]
\centering
\resizebox{0.6\hsize}{!}{\includegraphics{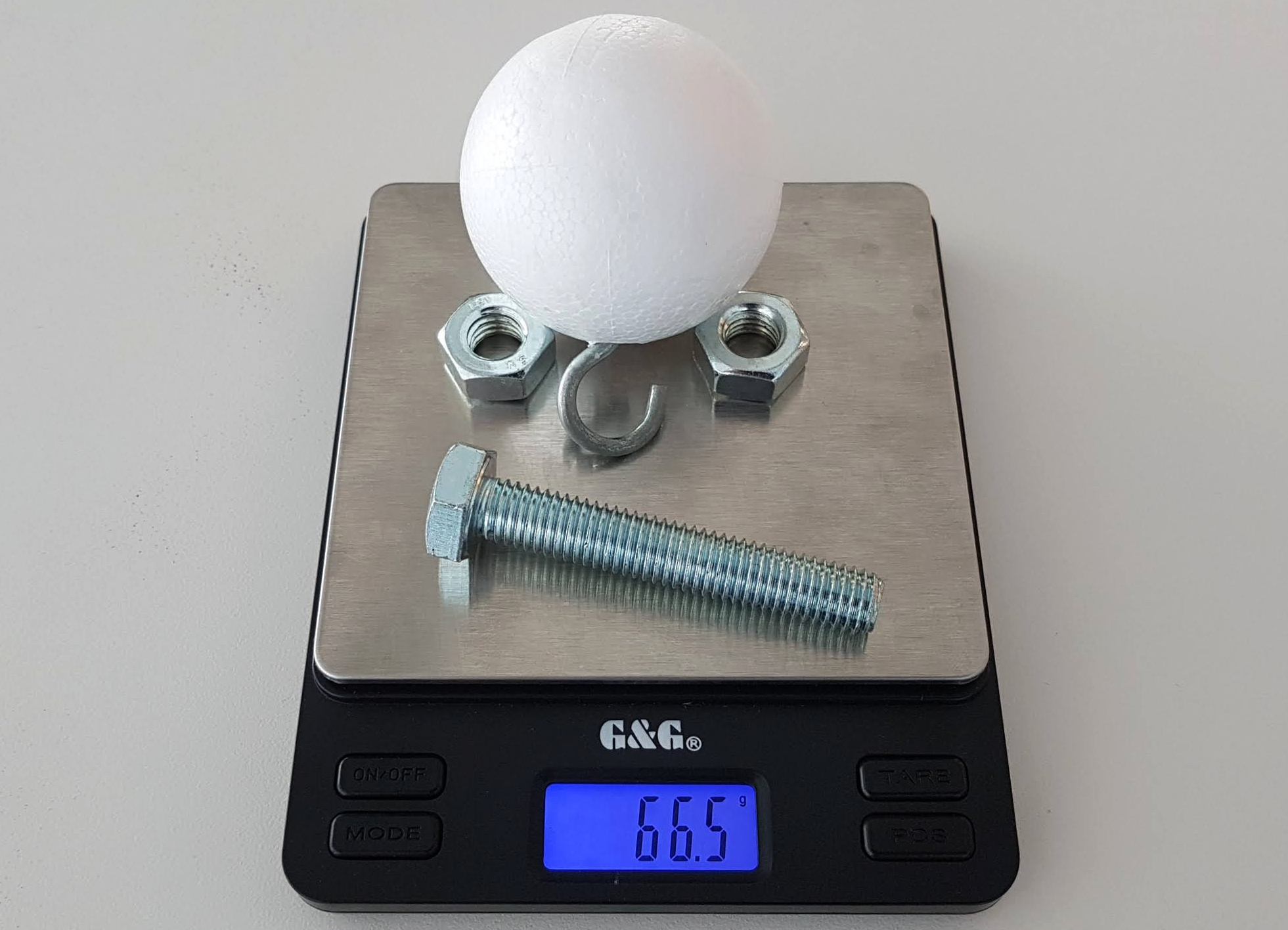}}
\caption{Das anfängliche Gewicht der Styroporkugel mit den entsprechenden Gewichten (Bild: M. Nielbock).}
\label{f:exp_step2}
\end{figure}

\medskip
In diesem Beispiel hat die Kugel einen Radius von \unit[2,5]{cm}. Damit hat sie ein Volumen von \unit[65,45]{cm$^3$} oder \unit[$6,545\cdot 10^{-5}$]{m$^3$}. Mit der Dichte des Wasser aus Tab.~\ref{t:groessen} folgt eine Masse von \unit[65,25]{g}. Zu Beginn sollte die Kugel also so beschwert werden, dass eine Gesamtmasse von etwa diesem Wert erreicht wird (Abb.~\ref{f:exp_step2}).

Dieses Gewicht wird nun mit Büroklammern am Haken der Kugel angebracht und im Wasserbecken eingetaucht.

\begin{figure}[!ht]
\centering
\resizebox{0.7\hsize}{!}{\includegraphics{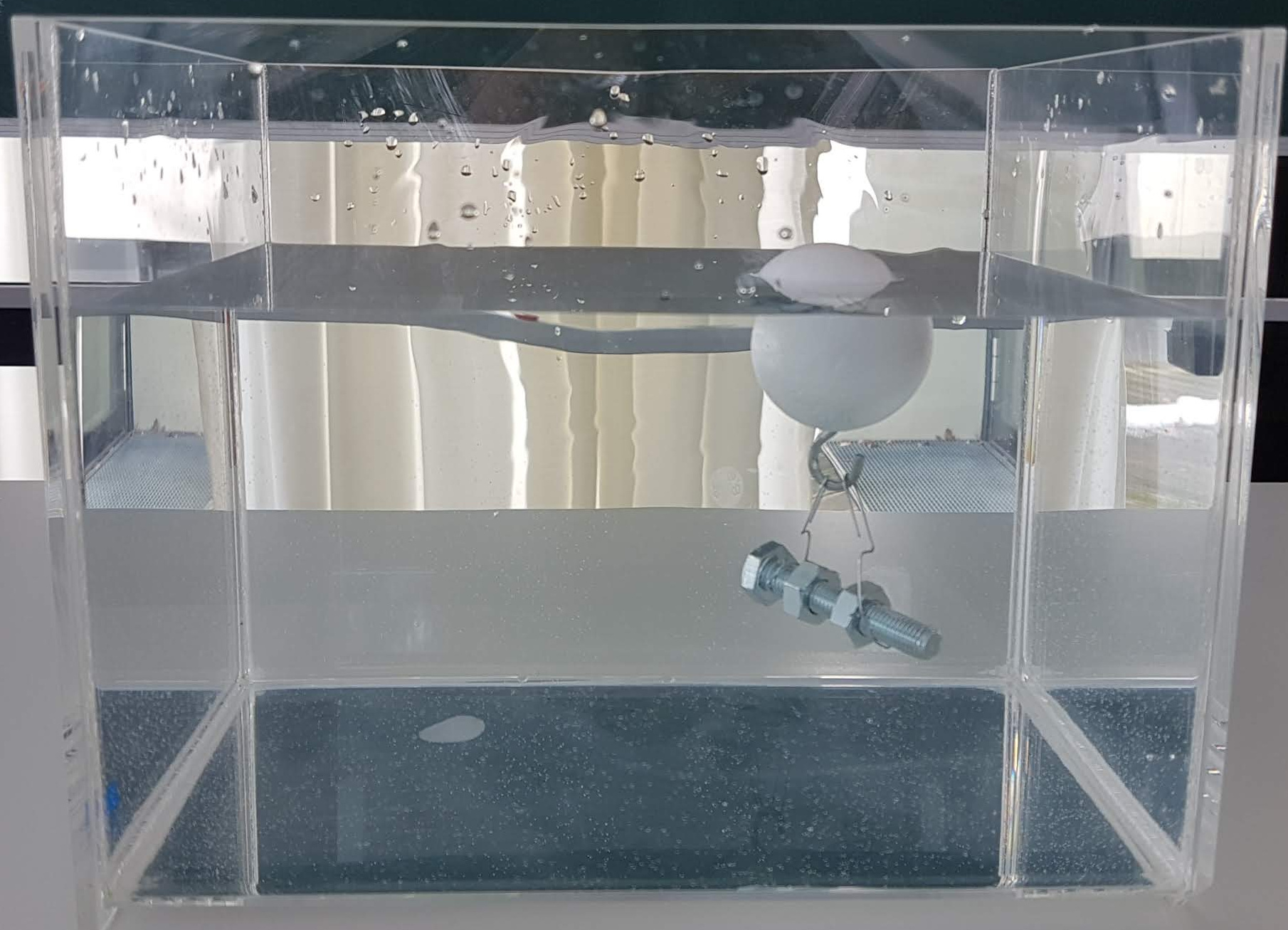}}
\caption{Das ermittelte Gesamtgewicht reicht nicht aus, um die Auftriebskraft auszugleichen. Die Kugel schwimmt teilweise eingetaucht immer noch an der Oberfläche (Bild: M. Nielbock).}
\label{f:exp_step3}
\end{figure}

\subsubsection*{Beschreibe und erkläre deine Beobachtung. Erkläre, warum die Kugel nicht völlig eingetaucht im Wasser schwimmt.}
Die Kugel schwimmt teilweise eingetaucht immer noch an der Oberfläche (Abb.~\ref{f:exp_step3}). Die benötigte Masse wurde offenbar unterschätzt. Der Grund ist, dass das Volumen der Gewichte vernachlässigt wurde. Auch sie erfahren einen Auftrieb.

\subsubsection*{Beschwere die Kugel, bis sie im Wasser schwebt.}
\begin{figure}[!ht]
\centering
\resizebox{0.7\hsize}{!}{\includegraphics{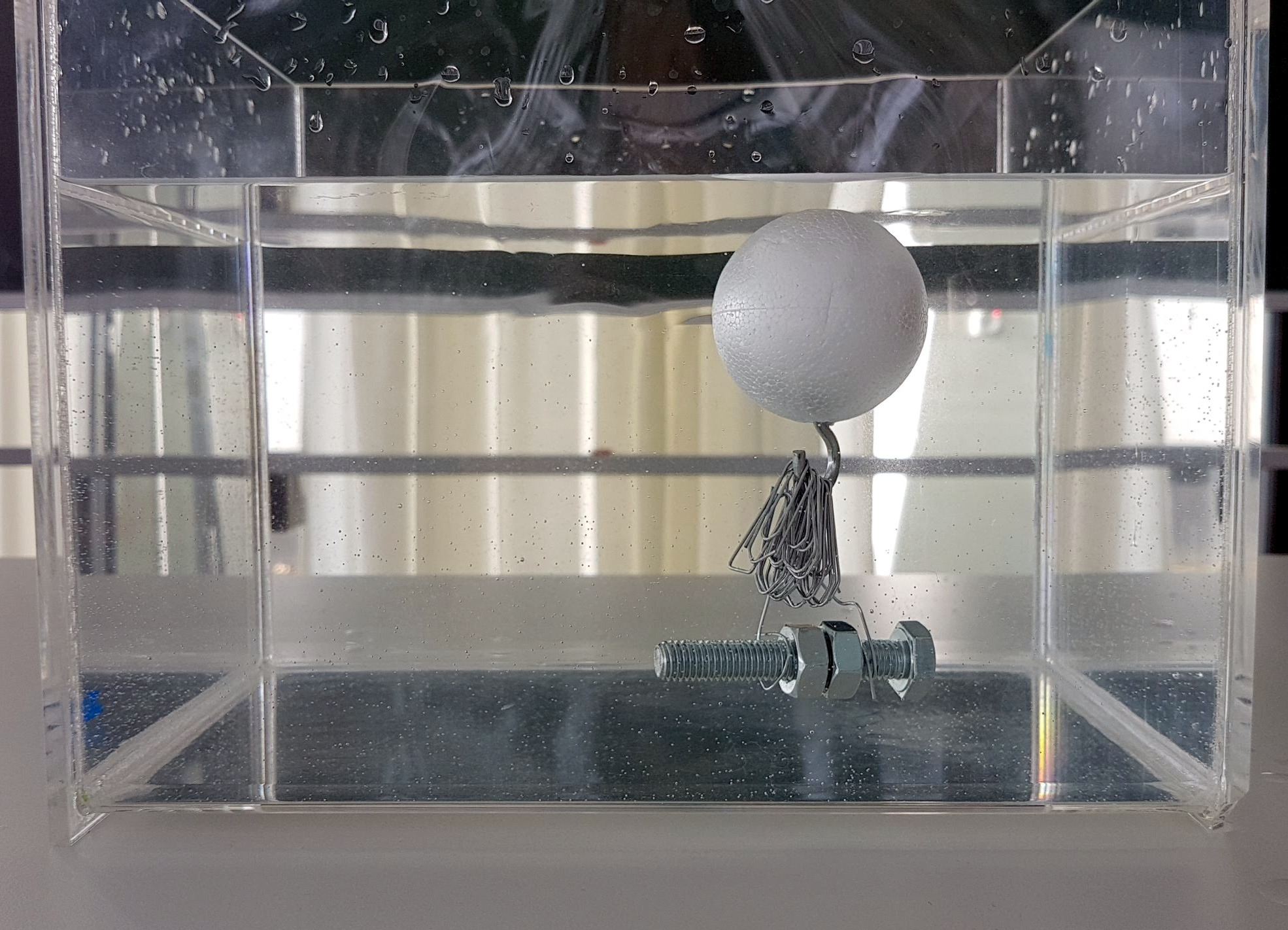}}
\caption{Nach der Zugabe einer ausreichenden Zahl von Büroklammern schwebt die Kugel im Wasser (Bild: M. Nielbock).}
\label{f:exp_step4}
\end{figure}

Nun beschwert man die Kugel so lange -- z.~B.~mit Büroklammern -- bis sich der Schwebezustand einstellt (Abb.~\ref{f:exp_step4}).

Dabei wird man feststellen, dass der Schwebezustand einem feinen Gleichgewicht zwischen Auftrieb und Gewicht entspricht. Unter Umständen ist die Gewichtseinheit pro Büroklammer zu grob, so dass die Kugel entweder langsam steigt oder sinkt.

\subsubsection*{Begründe, warum es so schwierig ist, eine exakte Balance aus Gewichtskraft und Auftrieb einzustellen.}

Jedes Ungleichgewicht führt sofort zu einer resultierenden Kraft in die eine oder andere Richtung. Da in diesem Experiment die kleinste Gewichtseinheit eine Büroklammer ist, können kleinere Gewichte nur dadurch erzielt werden, wenn man Stücke von einer Büroklammer entfernt. Hierzu wird die Zange benutzt.

\subsubsection*{Erkläre den  Zusammenhang besteht zwischen dem Volumen der Styroporkugel (plus Gewichte) und dem benötigten Gewicht.}

Der Zusammenhang zwischen dem Volumen der Kugel bzw. den Gewichten und dem benötigten Gewicht besteht in dem Archimedischen Prinzip. Der Auftrieb ist eine Kraft, die dem Gewicht des Wassers entspricht, das von dem eingetauchten Volumen verdrängt wird. Wenn man also weiß, wie groß das Volumen ist, entspricht die Auftriebskraft der Gewichtskraft dieses Wasservolumens. Daher benötigt man exakt so viel Gewicht zum Beschweren, wie das verdrängte Wasser wiegt.

\subsubsection*{Der Auftrieb ist in allen Medien vorhanden, so auch in der Luft. Schätze die Stärke des Auftriebs in der Luft im Vergleich zum Wasser ab. Nenne die maßgeblichen Größen.}

Das Verhältnis des Auftriebs in Luft und Wasser entspricht dem Dichteverhältnis dieser Medien. Mit der Dichte für Luft von $\varrho_L = \unit[1,2]{kg/m^3}$ findet man:

\begin{displaymath}
\frac{\varrho_w}{\varrho_L} = 831
\end{displaymath}

\subsubsection*{Optionale Zusatzaufgabe: Ermittle das während des Experiments verdrängte Volumen.}

\begin{figure}[!ht]
\centering
\resizebox{0.6\hsize}{!}{\includegraphics{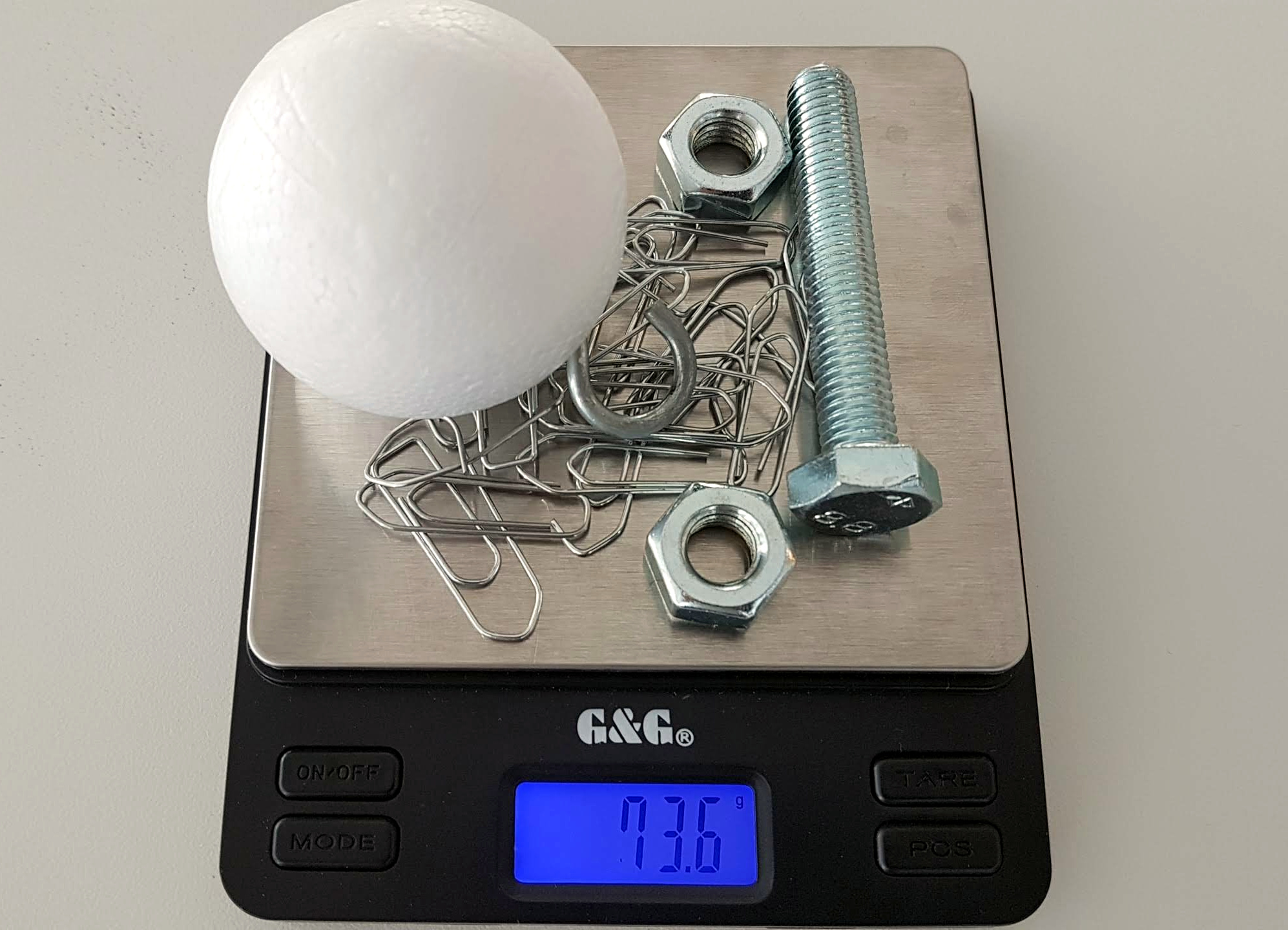}}
\caption{Die Waage zeigt eine Gesamtmasse von \unit[73,6]{g} an (Bild: M. Nielbock).}
\label{f:exp_step5}
\end{figure}

Auch hier lässt sich das Archimedische Prinzip anwenden. Aus der Gesamtmasse folgt über die Dichte des Wassers das verdrängte Volumen. Laut Waage entspricht das Gesamtgewicht einer Masse von \unit[73,6]{g} (Abb.~\ref{f:exp_step5}). Daraus folgt ein Volumen von \unit[73,8]{cm$^3$}.

\clearpage
\subsection*{Aufgaben}
Die nachfolgenden Aufgaben sollen dazu beitragen, die Inhalte durch Rechnungen weiter zu vertiefen.

\subsubsection*{1. Wasserdruck (rechnerisch)}

Der Wasserdruck nimmt proportional zur Tiefe zu. Gemäß Gl.~\ref{e:druck} lautet der Druck unter einer Wassersäule:

\begin{equation}
p = \varrho_w \cdot g \cdot h + p_0 \nonumber
\end{equation}

\medskip
Berechne die Wassertiefe, bei der ein Druck erreicht wird, der zusätzlich zur Atmosphäre über dem Wasser ebenfalls dem Normaldruck von $p-p_0 = p_0 = 101325\,\mathsf{Pa}$ entspricht. Die relevanten Zahlenwerte findest du in Tab.~\ref{t:groessen}.

\medskip
Beachte, dass gilt:

\begin{displaymath}
\unit[1]{Pa} = \unit[1]{\frac{N}{m^2}} = \unit[1]{\frac{\frac{kg\cdot m}{s^2}}{m^2}} = \unit[1]{\frac{kg}{m\cdot s^2}}
\end{displaymath}

\subsubsection*{2. Wasserdruck (grafisch)}

Die Gleichung aus Aufgabe 1 entspricht einer mathematischen Funktion der Form:

\begin{displaymath}
f(x) = a \cdot x + b
\end{displaymath}

\medskip
Es handelt sich dabei um eine Geradengleichung. Ordne die Größen denen der Gleichung aus Aufgabe 1 zu. Achte dabei darauf, bei welchen Größen es sich um Variablen handelt und welche davon konstant sind. Bestimme insbesondere die Steigung der Geraden $a$. 

\medskip
Konstruiere ein Koordinatensystem, das auf der senkrechten Achse den Druck angibt, während die horizontale Achse die variable Größe darstellt. Bestimme den Achsenabschnitt und wähle mit der Maßgabe, dass der Gesamtdruck das Doppelte von $p_0$ annehmen soll, die Skala der senkrechten Achse. Zeichne die Gerade. Ermittle den Wert auf der horizontalen Achse, für den $p = 2\cdot p_0$ beträgt.

\subsubsection*{3. Wasserverdrängung}

Berechne das Volumen eines Raumanzugs, der mit Astronaut eine Gesamtmasse von \unit[200]{kg} hat, wenn er im Wasser schwebt. Benutze das Gleichgewicht der Kräfte der Gravitation und des Auftriebs.

\subsubsection*{4. Das Archimedische Prinzip}

Das Archimedische Prinzip geht auf eine überlieferte Geschichte vom antiken griechischen Gelehrten Archimedes von Syrakus zurück. Demnach soll König Hieron II von Syrakus ihn beauftragt haben, einen möglichen Betrug zu untersuchen. Hieron hatte einem Goldschmied eine genau abgemessene Masse Gold für die Anfertigung einer Krone übergeben. Er vermutete aber, dass bei der Herstellung ein Teil des Golds durch Silber ersetzt wurde, wobei die Gesamtmasse aber erhalten blieb. Archimedes sollte nun herausfinden, ob der Verdacht gerechtfertigt war, und zwar ohne die Krone zu zerstören.

\medskip
Der römische Architekt Vitruv\endnote{Marcus Vitruvius Pollio war ein römischer Architekt, Ingenieur und Architekturtheoretiker des 1. Jahrhunderts v. d. Z.}, der später über diesen Fall berichtete, gab an, dass Archimedes dem Betrug dadurch auf die Schliche kam, weil eine Legierung von Gold und Silber eine geringere Dichte aufweist und deswegen durch sein höheres Volumen pro Masse auch mehr Wasser verdrängt. Als Archimedes dies durch Eintauchen in Wasser überprüfte, soll im Vergleich zu einem Stück reinem Gold bei der Krone das Wasser übergelaufen sein.

\medskip
Ist diese Geschichte glaubhaft? Berechne hierzu, wie groß der Unterschied zwischen den Wasserständen für zwei Metallstücke mit unterschiedlichem Goldanteil ist. Nimm hierzu zwei Stücke zu je \unit[1]{kg} an. Eines soll zu 100\% aus Gold bestehen; das andere soll eine der Masse nach 30\%ige Beimischung von Silber haben. Nimm zudem einen Behälter mit \unit[1]{$\ell$} Wasser an, dessen kreisrunde Grundfläche einen Radius von \unit[5]{cm} aufweist. Bestimme den Wasserstand,

\begin{enumerate}
    \item wenn der Behälter nur mit Wasser gefüllt ist,
    \item wenn das Goldstück vollständig eingetaucht ist,
    \item wenn das Stück aus der Gold-Silber-Legierung vollständig eingetaucht ist.
\end{enumerate}

Ermittle damit, ob der Unterschied der Wasserstände präzise genug messbar ist, um einen Unterschied zwischen den Metallstücken nachzuweisen.

\medskip
Alternativ wird angenommen, dass Archimedes stattdessen das verminderte Gewicht gemessen hat, das der Auftrieb im Wasser erzeugt. Berechne dazu die Auftriebskräfte für die beiden Metallstücke und deren Differenz, wenn sie in Wasser eingetaucht sind. Bestimme die Masse, die dieser Kraftdifferenz entspricht.

\clearpage
\subsection*{Lösungen}

\subsubsection*{1. Wasserdruck (rechnerisch)}

\begin{equation}
h = \frac{p -p_0}{\varrho_w \cdot g} = \frac{101325\,\mathsf{Pa}}{997\,\mathsf{kg/m}^3 \cdot 9,81\,\mathsf{m/s}^2} = 10,4\,\mathsf{m} \nonumber
\end{equation}

\subsubsection*{2. Wasserduck (grafisch)}

Die Geradengleichung lautet:

\begin{displaymath}
p(h) = \left(\varrho_w \cdot g\right) \cdot h + p_0
\end{displaymath}

\medskip
Steigung: $\varrho_w \cdot g = \unit[997]{\frac{kg}{m^3} \cdot \unit[9,81]{\frac{m}{s^2}} = \unit[9780,57]{\frac{kg}{m^2\cdot s^2}}}$

\medskip
Achsenabschnitt: $p_0 = \unit[101325]{Pa} = \unit[101325]{\frac{kg}{m\cdot s^2}}$

\medskip
Für $p(h) = 2\cdot p_0$ muss gelten:

\begin{displaymath}
\unit[9780,57]{\frac{kg}{m^2\cdot s^2}} \cdot h = p_0 = \unit[101325]{\frac{kg}{m\cdot s^2}}
\end{displaymath}

\begin{figure}[!ht]
\centering
\resizebox{\hsize}{!}{\includegraphics{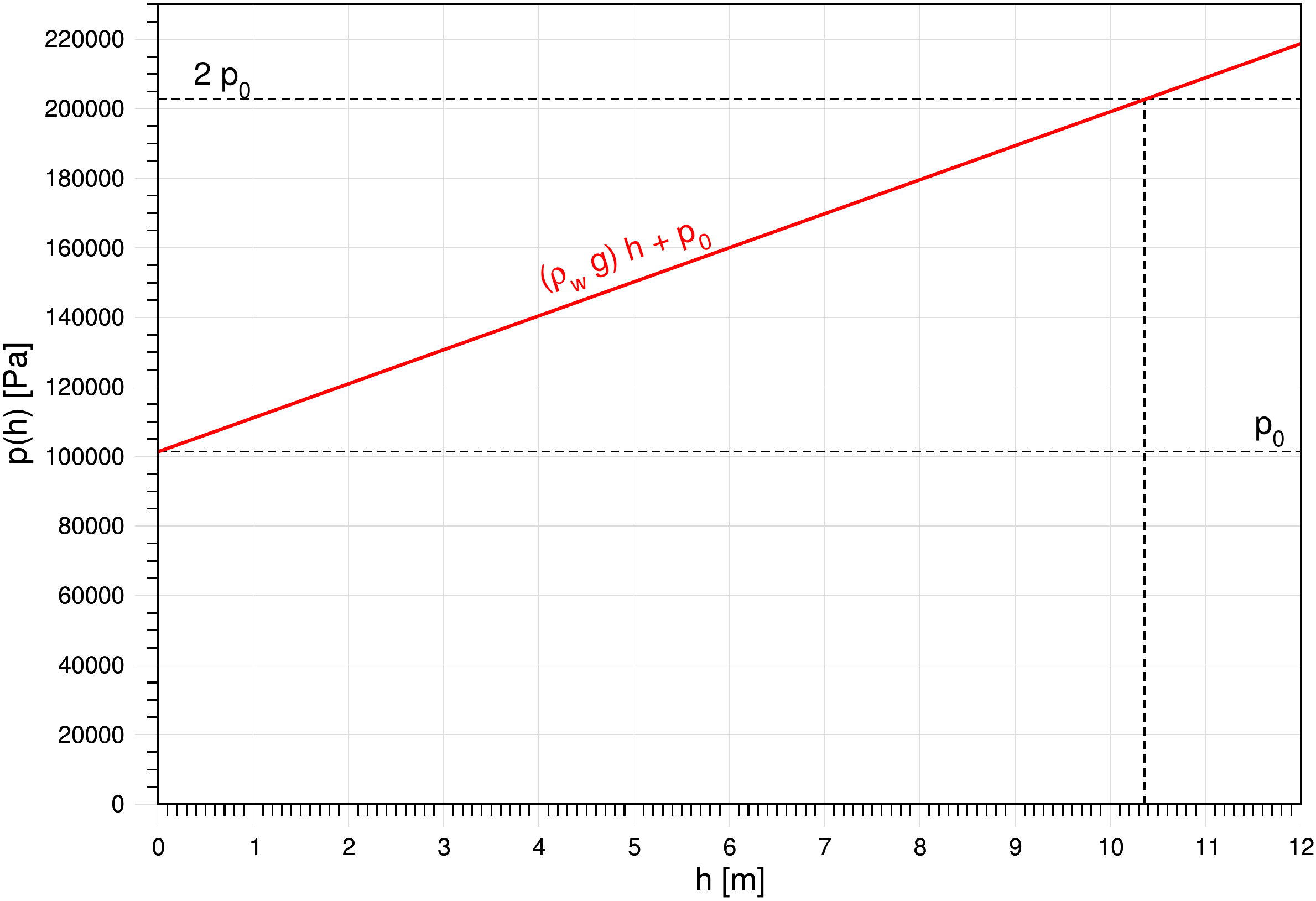}}
\label{f:graph}
\end{figure}

\subsubsection*{3. Wasserverdrängung}

Wenn der Astronaut mit Raumanzug im Wasser schwebt, stehen Schwerkraft und Auftriebskraft im Gleichgewicht. Gemäß Gl.~\ref{e:Fgg} gilt:

\begin{align}
    F_g &= F_a \notag \\[5pt]
    \Leftrightarrow m\cdot g &= m_w \cdot g \notag \\[5pt]
    \Leftrightarrow m &= \varrho_w \cdot V \notag \\[5pt]
    \Leftrightarrow V &= \frac{m}{\varrho_w} \notag \\[5pt]
                      &= \frac{200\,\mathsf{kg}}{997\,\mathsf{kg/m}^3} = 0,2\,\mathsf{m}^3 \notag
\end{align}

\subsubsection*{4. Das Archimedische Prinzip}

Aus den Werten in Tab.~\ref{t:groessen} kann man die Dichte der Legierung berechnen, die einen Silberanteil von 30\% hat.
\begin{equation}
    \varrho_{70/30} = 0,7\cdot\varrho_\mathsf{Au} + 0,3\cdot\varrho_\mathsf{Ag} = 1672\,\mathsf{kg/m}^3
\end{equation}

\medskip
Metallstücke aus Gold, Silber und der Legierung von je \unit[1]{kg} besitzen die folgenden Volumina.

\begin{align}
    V_\mathsf{Au} &= \frac{m}{\varrho_\mathsf{Au}} = \frac{1000\,\mathsf{kg}}{1939\,\mathsf{kg/m}^3} = 0,5157\,\mathsf{m}^3 \notag \\[5pt]
    V_\mathsf{Ag} &= \frac{m}{\varrho_\mathsf{Ag}} = \frac{1000\,\mathsf{kg}}{1049\,\mathsf{kg/m}^3} = 0,9533\,\mathsf{m}^3 \notag \\[5pt]
    V_{70/30}     &= \frac{m}{\varrho_{70/30}}     = \frac{1000\,\mathsf{kg}}{1672\,\mathsf{kg/m}^3} = 0,5981\,\mathsf{m}^3 \notag
\end{align}

Im relativen Vergleich $\frac{V_{70/30}}{V_\mathsf{Au}} = 1,16$ scheint der Unterschied sehr wohl messbar zu sein. Für ein Wasservolumen von \unit[25]{$\ell$} in einem Becher mit dem Radius \unit[25]{cm} erhält man folgende Wasserstände.

$$h=\frac{V}{A}=\frac{V}{\pi\cdot r^2}$$

\begin{enumerate}
    \item $V_w = 25000\,\mathsf{cm}^3 \Rightarrow h = 12,73\,\mathsf{cm}$
    \item $V_w+V_\mathsf{Au} = 25051,57\,\mathsf{cm}^3 \Rightarrow h = 12,76\,\mathsf{cm}$
    \item $V_w+V_{70/30} = 25059,81\,\mathsf{cm}^3 \Rightarrow h = 12,76\,\mathsf{cm}$
\end{enumerate}

Der Wasserstand verändert sich kaum. Daher ist es unwahrscheinlich, dass Archimedes den Betrug auf diese Weise aufgedeckt hat.

\medskip
Die Kraft des Auftriebs ist definiert als:

$$F_a = \varrho_w\cdot V \cdot g = \varrho_w\cdot \frac{m}{\varrho_\mathsf{Metall}} \cdot g = \frac{\varrho_w}{\varrho_\mathsf{Metall}} \cdot m \cdot g$$

\medskip
Somit erhält man für das Verhältnis der Auftriebskräfte:

\begin{equation*}
    \frac{F_{a,70/30}}{F_{a,\mathsf{Au}}} = \frac{V_{70/30}}{V_\mathsf{Au}} = 1,16
\end{equation*}

\medskip
Die Auftriebskräfte für \unit[1]{kg} Gold und \unit[1]{kg} der Legierung lauten:

\begin{align*}
F_{a,70/30}       &= \frac{997\,\mathsf{kg/m}^3}{1672\,\mathsf{kg/m}^3} \cdot 1\,\mathsf{kg} \cdot 9,81\,\mathsf{m/s}^2 = 5,850\,\mathsf{N}\\[5pt]
F_{a,\mathsf{Au}} &= \frac{997\,\mathsf{kg/m}^3}{1939\,\mathsf{kg/m}^3} \cdot 1\,\mathsf{kg} \cdot 9,81\,\mathsf{m/s}^2 = 5,044\,\mathsf{N}
\end{align*}

\medskip
Die Differenz beträgt $F_{a,70/30} - F_{a,\mathsf{Au}} = 0,806\,\mathsf{N}$. Das entspricht einer Masse von \unit[82]{g}. Dies war auch zu Zeiten von Archimedes gut messbar.

\clearpage
 { \parindent 0pt
     \parskip 2ex
    \def\enotesize{\normalsize}
     \theendnotes   }


\printbibliography

\clearpage
\section*{Danksagung}
Der Autor bedankt sich bei den Lehrern Matthias Penselin, Florian Seitz und Martin Wetz für ihre wertvollen Hinweise, Kommentare und Änderungsvorschläge, die in die Erstellung dieses Materials eingeflossen sind. Weiterer Dank gilt Herrn Dr. Volker Kratzenberg-Annies für seine gewissenhafte Durchsicht.

\vfill
\medskip
Diese Unterrichtsmaterialien sind im Rahmen des Projekts {\em Raum für Bildung} am Haus der Astronomie in Heidelberg entstanden. Experimente dürfen nur unter Aufsicht einer Lehrkraft durchgeführt werden. Der Autor und die Projektpartner übernehmen keine Haftung für etwaig auftretende Verletzungen. Weitere Materialien des Projekts finden Sie unter:

\begin{center}
\href{http://www.haus-der-astronomie.de/raum-fuer-bildung}{http://www.haus-der-astronomie.de/raum-fuer-bildung}
und
\href{http://www.dlr.de/next}{http://www.dlr.de/next}
\end{center}

Das Projekt findet in Kooperation mit dem Deutschen Zentrum für Luft- und Raumfahrt statt und wird von der Joachim Herz Stiftung gefördert.

\medskip
\begin{center}
\includegraphics[height=1.5cm]{HdA_500x260.png}
\hspace*{4em}
\includegraphics[height=1.5cm]{DLR_Logo_svg.png}
\hspace*{4em}
\includegraphics[height=1.5cm]{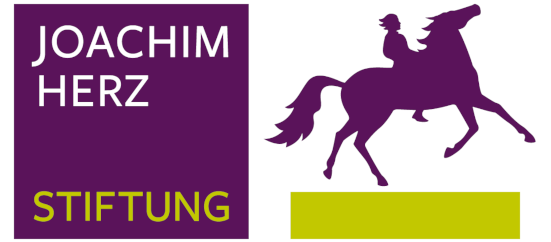}
\end{center}

\end{document}